\documentclass[12pt]{article}
\usepackage{amsmath}
\usepackage{graphicx}
\usepackage{amsthm}
\usepackage{dsfont}
\usepackage{amssymb}
\usepackage{subcaption}
\usepackage{dsfont}
\usepackage{url}

\newcommand{\beqn}{\begin{eqnarray*}}
\newcommand{\eeqn}{\end{eqnarray*}}
\newcommand{\bneqn}{\begin{eqnarray}}
\newcommand{\eneqn}{\end{eqnarray}}
\newcommand{\parens}[1]{\left(#1\right)}

\newcommand{\bracks}[1]{\left[#1\right]}

\newcommand{\expe}[1]{\mathbb{E}\bracks{#1}}

\newcommand{\braces}[1]{\left\{#1\right\}}
\newcommand{\prob}[1]{\mathbb{P}\parens{#1}}

\usepackage{bbm}
\usepackage{color}
\usepackage{booktabs}
\usepackage{multirow}
\usepackage{setspace}
\usepackage{caption}
\def\pconst{\nu}

\def\pconst{\nu}

\title{Thick distribution tails in models of cancer secondary tumors}
\author{Philip A.Ernst$^\text{a}$, Marek Kimmel$^{\text{a,b,c}}$, Monika Kurpas$^\text{c}$, and Quan Zhou$^\text{a}$}
\date\today

\begin{document}
\maketitle

\begin{abstract}
Recent progress in microdissection and in DNA sequencing has enabled subsampling of multi-focal cancers in organs such as the liver in several hundred spots, helping to determine the pattern of mutations in each of these spots. This has led to the construction of genealogies of the primary, secondary, tertiary and so forth, foci of the tumor. These studies have led to diverse conclusions concerning the Darwinian (selective) or neutral evolution in cancer. Mathematical models of development of multifocal tumors have been developed to support these claims. We report a model of development of a multifocal tumor, which is a mathematically rigorous refinement of a model of \cite{Ling}. Guided by numerical studies and simulations, we show that the rigorous model, in the form of an infinite-type branching process, displays distributions of tumors size which
have heavy tails and moments that become infinite in finite time. To demonstrate these points, we obtain bounds on the tails of the distributions of the process and infinite-series expression for the first moments. In addition to its inherent mathematical interest, the model is corroborated by recent reports of apparent super-exponential growth in cancer metastases.
\end{abstract}


\footnotetext[1]{Department of Statistics, Rice University, Houston, TX, USA}
\footnotetext[2]{Department of Bioengineering, Rice University, Houston, TX, USA}
\footnotetext[3]{Systems Engineering Group, Silesian University of Technology, Gliwice, Poland}

\section{Introduction}

Growth patterns and heterogeneity of cancer metastases are not well
understood, although it seems clear that they are a product of mutation,
genetic drift, migration and selection and perhaps other population
genetics and population dynamics mechanisms. Recently, it was
reported by \cite{Barat} that in some animal models,
metastases exhibit growth pattern that appears to be super-exponential.
In a series of careful experiments and computations, the authors provided
an explanation which involved rather complicated biological mechanisms.
In this paper, we attempt to demonstrate that similar behavior may be
generated by a very simple growth and mutation model.

Our work is also motivated by the recent work \cite{Ling}, which presents an
analysis of a sequencing experiment using the nearly 300 samples taken
from a section of hepatocellular carcinoma tumor. The purpose of the
analysis has been to compare the Darwinian and non-Darwinian theories
of development of human solid cancers. Their Darwinian model involves a primary
tumor growing and shedding secondary foci with distributed growth
rates, which seems an attractive way of modeling competition among
the secondary foci. However, when examining the details of their approach,
we found that the ``Darwinian\textquotedblright{} model introduced
by \cite{Ling} (and also treated in \cite{Tao}) displays
a peculiar behavior, characterized by existence of outlier trajectories
and explosion of expected cell count in finite time. We trace this
behavior to the exponential model used by the authors as a distribution
of growth rates.

A very similar behavior is exhibited by a simple ``toy model'' that
involves exponential growth curve with Malthusian parameter (growth
rate) that itself is an exponentially distributed random variable.
In the toy model the explosions are related to the distributions of
population size being of Pareto type with coefficients changing in
time. However, none of these models (neither the model of \cite{Ling} nor the toy model) are truly stochastic in the sense
that they describe cell divisions and mutations as stochastic events
occurring as the cell population is evolving in time. This led us
to the idea of building a stochastic population model being a branching
process, in which for simplicity lifelengths of cells are assumed
to be exponentially distributed, and in which at each division one
progeny cell may mutate and acquire a new lifelength distribution
which is exponential with the parameter sampled from an exponential distribution.
This process may be classified as an age-dependent Markov branching
process with a non-denumerable type space. We show that the process exhibits
finite-time explosions of expected values, while simulations indicate
Pareto-like tails, with exponents changing in time and becoming equal
to 1 at the time the expectations explode. We develop a set of bounds that are consistent with the simulation findings.
We also prove the finite-time explosion of expected values of the process.

\section{Baseline Model}
This is in principle the model considered by \cite{Ling,Tao} (Figure~\ref{fig:fig1} (A)).
A primary tumor is generated from a single cell at time $t=0$ and grows at a rate $g(x)=bx$, where $x$ denotes the number of cells in the tumor and $b$ is a constant. The growing tumor emits transformed single cells at the rate $\beta(x)$, where $\beta(x)=mx^{\alpha}$. 
Initially, the constant $\alpha$ is set equal to $2/3$ to symbolize the fact that tumors shed new cells via their surface.
Each transformed cell develops into a new tumor, which grows at a generally different rate $g(x)$ and emits newly transformed cells just as the primary does, that is, $g(x)=ax$. In general, we will use $a$ and $b$ to denote the growth rate of a tumor, be it a primary tumor or a secondary tumor.

Following \cite{Iwata}, the dynamics of the secondary cell colony size distribution density are given by the following von Foerster-type equations
\beqn
\frac{\partial \rho(x,t)}{\partial t}+\frac{\partial g(x)\rho(x,t)}{\partial x}&=&0, \quad (x,t) \in [1,\infty) \times [0, \infty)\label{Baseline}\\
\rho(x,0)&=&0,
\eeqn
with nonlocal boundary conditions
\beqn
g(1)\rho(1,t)=\int_1^{e^{at}} \beta(x)\rho(x,t)dx+\beta(x_p(t)),\quad \beta(x)=mx^{\alpha}.\label{nonlocal}
\eeqn

As demonstrated in the Appendix, if growth rates $a$ and $b$ are constant, the solution has the form
\beqn
\rho(x,t)= \frac{m}{a(a\alpha+m-\alpha b)}((a\alpha-\alpha b)e^{\alpha bt}x^{-(\frac{\alpha b}{a}+1)}+me^{(a\alpha+m)t}x^{-(\alpha+\frac{m}{a}+1)}),
\eeqn
for $x<e^{at}$ and $\rho(x,t)=0$ otherwise. The tail distribution corresponding to the density $\rho(x,t)$ has the form
\bneqn
G(x)&=&G(x;a,b)=\int_{x}^{e^{at}}\rho(\xi,t)d\xi \\
& = & \frac{m}{a\alpha+m-\alpha b}(\frac{a-b}{b}(e^{\alpha bt}x^{-\frac{\alpha b}{a}}-1)+\frac{m}{a\alpha+m}(e^{(a\alpha+m)t}x^{-\alpha-\frac{m}{a}}-1)),
\nonumber 
\eneqn
for $x\in[1,e^{at}]$, and $G(x)=0$ for $x>e^{at}$. This function increases exponentially with rate $\max(\alpha b,\alpha a+m)$. 
However, if the growth rate of secondary tumors $a$ is a random variable with exponential distribution, 
then
\beqn
\tilde{G}(x;b)&=&\int_{ln(x)/t}^\infty G(x;a,b)\lambda \text{exp}(-\lambda a)da.
\eeqn
However, substitution of the expression for $G(x;a,b)$ leads to intractable integrals, except for the case $x=1$ (total count of secondary foci), when it leads to the following expression
\begin{multline*}
\tilde{G}(1;b)=\frac{m(e^{b\alpha t}-1)}{b\alpha}+(2-e^{b\alpha t})\frac{m^2\lambda}{b\alpha^2}e^{\lambda(m/\alpha-b)}\Gamma(0, \lambda(m/\alpha-b))+\\
\frac{m^2\lambda}{b\alpha^2}\braces{-e^{mt}\bracks{e^{(\lambda-\alpha t)(m/\alpha-b)}\Gamma(0, (\lambda-\alpha t)(m/\alpha-b))-e^{(\lambda-\alpha t)(m/\alpha)}\Gamma(0, (\lambda-\alpha t)m/\alpha}-e^{\lambda m/\alpha}\Gamma(0, \lambda m/\alpha)},
\end{multline*}
where $\Gamma(z,w)=\int_w^\infty e^{-t}t^{z-1}dt$ is the incomplete Gamma function, 
which behaves as $-\text{log}(w)$ as $w \downarrow 0$. 
Thus the solution increases to infinity as $t\uparrow\lambda/\alpha$. This highly irregular behavior of the ``quasistochastic" version of 
the Baseline Model inspired us to seek a fully stochastic model with analogous behavior. We present this model in the following section.

\section{Branching process model}
\subsection{Stochastic toy model}
Does a truly stochastic model display the same value behavior? Let us try a toy model, as follows. Let secondary tumors grow exponentially at rate $a$, which itself is a random variable, as follows
\beqn
X(t\mid a)=\text{exp}(at), \quad t\geq 0, \quad a \sim \text{exp}(\lambda).
\eeqn
It now has Pareto tail
\beqn
\prob{X(t)>x}= \begin{cases}
1, \quad 0 \leq x <1,\\
x^{-\lambda/t}, \quad x \geq 1,
\end{cases}
\eeqn
for $t \geq 0$.
We integrate the above to obtain:
\beqn
\expe{X(t)}=\int_0^\infty \prob{X(t)>x}dx=1+\int_1^\infty x^{-\lambda/t}dx=\begin{cases}
\lambda(\lambda-t)^{-1}, \quad t<\lambda\\ \infty, \quad t \geq \lambda 
\end{cases}.
\eeqn
What about the $\infty$? Will tumors really \textit{explode}? We now turn our attention to this matter.

\subsection{Modified Goldie-Coldman Model} \label{sec32}
We now consider a modified Goldie-Coldman (G-C) model (Figure~\ref{fig:fig1} (B)). The classical version can be found in \cite{Kimmel}.
\begin{enumerate}
 \item Cells are organized in proliferating clones characterized by division rates $a$.
Within each clone, cells proliferate according to a time-continuous Markov branching process 
with perfect binary fission and usual independence assumptions, i.e., their 
life-lengths are exponentially distributed
with parameter $a$. No cell death is considered. Cell type is identical
with its division rate.
\item At any division, with probability $\mu$, one cell mutates and assumes division rate sampled from exponential distribution with parameter $\lambda$ ($a'\sim \text{exp}(\lambda)$). 
\item The process is started by an ancestor cell with a fixed type $a$.
In a variant of the process, the ancestor cell type is sampled from exponential distribution 
with parameter $\lambda$ ($a\sim \text{exp}(\lambda)$).
\end{enumerate}
The resulting  model is a continuum-type time-continuous Markov branching process.
An ODE can be written for the probability generating function (pgf) of the distribution of total cell counts in all clones.

We start from presenting simulation results, which motivate the more mathematical study that follows. We then perform some asymptotic calculations to characterize the tail distribution of the cell counts of different types. 
In particular, we will show that the tail probability of these cell counts can be bounded from below by a power law with exponent $-\lambda/(1 - \mu) t$. \\
\indent Formally, consider one ancestor tumor cell with division rate $a$ at time $0$. 
At each division, with probability $\mu$, it can divide into one cell with rate $a$ and another cell with division rate $a'$ where $a' \sim \text{exp}(\lambda)$. 
The new type of tumor cells (with rate $a'$) have the same mutation rate, $\mu$, and can continue mutating into new subtypes with random division rate generated from $\text{exp}(\lambda)$. 
All the tumor cells are assumed to be independent of one another. 
Let $X_k(a, t)$ be the number of tumor cells that are generated by $k - 1$ mutations.  Accordingly, $X_1(a, t)$ denotes the number of primary tumor cells, i.e., the cells with division rate $a$;
$X_2(a, t)$ denotes the number of cells of types that directly mutated from primary tumor cells; 
$X_3(a, t), X_4(a, t), \dots, $ are defined analogously. 

Finally, we derive the equations for the probability generating functions of the total count of cells in the process. 
We proceed to derive an infinite series solution for the expected counts of cells and show that it explodes in finite time,
almost exactly as it does in the toy model.

\subsection{Simulation results}

We begin with the toy model, which provides guidance concerning the
behavior of the branching process model. For the version of the toy
model with $\lambda=1$, Figure~\ref{fig:fig1} (C) depicts the true expected value
$\expe{X(t)}$ of the process (which explodes at $t=1$), averages of
1000 realizations of $X(t)$, and 0.5 and 0.95 quantiles of $X(t)$,
all in semi-logarithmic scale. Notice that the averages increase
faster than any exponential, while the quantiles grow exponentially.
The explosion at $t=1$ is analogous to the behavior exhibited by
the baseline model.

We turn to the Modified G-C Model. We carried out extensive simulations
of the model, assuming widely ranging parameters. Selected results
are depicted in Figure~\ref{fig:fig1} (D) and 
Figures~\ref{fig:fig2} -- \ref{fig:fig5}. Figure~\ref{fig:fig1} (D) is based on 10,000 simulated
trajectories of the Modified G-C process with parameters $\mu=.5, a=.01, \lambda=10$.  Depicted are realizations of the process
ranking 1-10 (green), 51-100 (red), and 301-400 (blue) at time $t=20$.
The distribution of trajectories exhibits strong right skewness and
suggests heavy tails. Figures~\ref{fig:fig2} -- \ref{fig:fig4} depict averages of the simulated
trajectories of the Modified G-C process with three different cases: $\mu=0.5, a=0.01, \lambda=10$ (Fig. 2),
$\mu=0.5, a=0.01, \lambda=100$ (Fig. 3), and
$\mu=0.1, \lambda=100$ (Fig. 4), based on 200, 1000, and 10,000 trajectories,
with the expectations $M(a,t)$, computed by numerically solving the
integral equation (\ref{integraleq}) for $\varphi(t)$ and using expression
(\ref{ConvolEqu}). The averages are convex in semi-log coordinates, which
suggests faster than exponential growth. However, they underestimate
the growth of the expectation, which explodes to infinity at $t=\lambda/(1-\mu)$. \\
\indent Figure~\ref{fig:fig5} depicts simulated tail behavior of the Modified G-C process.
Estimated power exponents of the tail of $X(a,t)$, approach value
-1 as $t\uparrow\lambda/(1-\mu)$ and examples of empirical tail in
log-log coordinates, approximated by a straight line. Due to the heavy tails
of $X(a,t)$, power exponents
based on simulations are underestimates. However, they seem to indicate that the expectation
of $X(a,t)$ tends to infinity as $t\uparrow\lambda/(1-\mu)$. This
intuition will be confirmed formally in the sequel.

\subsection{Asymptotic bounds}
In this section, we consider the distributions of $X_1(a, t)$, $X_2(a, t)$, and, in general, $X_k(a, t)$.

\subsubsection{Distribution of $X_1(a, t)$}
Due to the independence assumption, the distribution of the primary tumor cells, $X_1(a, t)$, is not affected by the behavior of subtypes that mutated from the primary type. 
Standard results for Yule's binary fission model gives that 
\begin{align*}
F_1 (s, a, t ) =  \dfrac{s e^{-a (1 - \mu) t} }{1 - s (1 - e^{-a (1 - \mu) t})  },  \quad \quad  s \in [0, 1] ,  \;  t \geq 0, 
\end{align*}
where $F_1(s, a, t)$ is the probability generating function of $X_1(a, t)$. 
This is a geometric distribution with success probability $e^{-a(1 - \mu t)}$. Hence, 
\begin{equation}\label{eq:distr.x1}
\expe{ X_1(a, t) } = e^{a (1 - \mu) t} ,  \quad \quad  \prob{X_1(a, t) > n} = (1 - e^{- a(1 - \mu) t})^n . 
\end{equation}

We next introduce a result that will be very useful for studying the distribution of $X_2(a, t), X_3(a, t), \dots$. 
If we integrate over $a \sim \text{exp}(\lambda)$, the marginal distribution of $X_1$ is 
known as Yule-Simon distribution~\cite{simon1955class, yule1925mathematical}.  
Define 
\begin{equation}\label{eq:def.nu}
\pconst(t) \equiv \dfrac{\lambda }{(1 - \mu) t}. 
\end{equation}
The probability mass function and the tail probability of $X_1(t)$ are given by 
\begin{equation}\label{eq:z1.pmf}
\mathbb{P}(X_1(t) =  n)  =  \pconst \mathrm{B}(\pconst + 1, n ), \quad \quad 
\mathbb{P}(X_1(t) > n) =  n  \mathrm{B}(\pconst + 1, n ), 
\end{equation}
where B stands for the beta function. 
Note that for sufficiently large $n$, the tail probability follows a power law
\begin{equation}\label{eq:z1.tail}
\mathbb{P}(X_1(t) > n) =   \dfrac{ \Gamma(n + 1)  \Gamma (\pconst + 1)}{\Gamma (\pconst + n + 1) }  \sim  \dfrac{\Gamma(\pconst + 1)}{n^\pconst }, \quad \quad n \rightarrow \infty. 
\end{equation} 
The first two moments of $X_1(t)$ are 
\begin{equation}\label{eq:z1.moments}
\begin{aligned}
\expe{  X_1(t) } = \left\{ \begin{array}{cc}
\dfrac{\pconst }{\pconst - 1}   &       \text{ if }  \pconst > 1, \vspace{0.2cm} \\
\infty    &       \text{ if }  \pconst \leq 1 ,    
\end{array}  \right. \quad \quad 
\mathrm{Var}(  X_1(t) ) = \left\{ \begin{array}{cc}
\dfrac{ \pconst^2 }{ ( \pconst - 1 )^2  ( \pconst - 2 ) }    &       \text{ if }  \pconst > 2 ,  \vspace{0.2cm} \\
\infty    &       \text{ if }   \pconst \leq 2 . 
\end{array}  \right.
\end{aligned}
\end{equation}
This is essentially the same as the result we obtained for the toy model introduced at the beginning of this section. 
\subsubsection{Distribution of $X_2(a, t)$}
Let $K(a, t)$ denote the number of tumor types generated by one and only one mutation.
Denote the division rates of these subtypes by $a'_1, \dots, a'_{K(a,t)}$ and let $Y_i(a, t)$ be the number of cells of type $a'_i$. Thus $X_2(a, t) = \sum_{i=1}^{K(a,t)} Y_i(a, t)$. 
Recall that $a$ is just the division rate of the ancestor tumor cell.
Hence the notation $Y_i(a, t)$ implies that $a'_i$ is integrated out. 
Clearly, if a subtype $a'_i$ is born at time $t_0 < t$, the distribution of $Y_i(a, t)$ is the same as the marginal distribution of $X_1(t - t_0)$. 
We can compute the expected value of $X_2(a, t)$ as
\begin{equation}\label{eq:z22.expe}
\expe{ X_2(a, t)  }  = \left\{  \begin{array}{cc}
\displaystyle\int_0^t   \dfrac{  \lambda a \mu e^{a_1 (1 - \mu) s} }{\lambda - (1-\mu)(t - s) } ds ,  &  (1 - \mu)t < \lambda , \vspace{0.2cm} \\
\infty,  &  (1 - \mu)t \geq \lambda.
\end{array}
\right.
\end{equation}

We now consider the tail probabilities $\mathbb{P}( X_{2}(a, t) > n)$, which may be bounded by 
\begin{equation}\label{eq:approx}
\mathbb{P}( X_2(a, t) > n)  =  \mathbb{P} \left(  \sum\limits_{i=1}^{K(a, t)} Y_i(a, t) > n \right) 
\geq   \mathbb{P}\left(  \bigcup\limits_{i=1}^{K(a, t)}  \{  Y_i(a, t)  > n  \}   \right) . 
\end{equation}
We pause to comment on why this bound could be useful. 
For a tumor model, $a$ is typically small and $\lambda$ is large so that the primary tumor type and most secondary tumor types do not grow too quickly. 
The mutation rate $\mu$ also takes a small value due to its biological meaning. 
Since, by \eqref{eq:z22.expe}, eventually the number of tumor cells will explode, our primary interest is in the case where $t$ is moderate, and consequently the event $\{ K(a, t) \geq 2\}$ has a  small probability. 
But a more important reason is that the tail probability of $Y_i$ is a power law. 
Thus, we are much more likely to observe one very large $Y_i$ than to observe two or more ``moderately large" $Y_i$'s.  
The left-hand side of~\eqref{eq:approx} can be computed as
\begin{equation}\label{eq:pn}
 P_n(a, t)  \equiv  \mathbb{P}\left(  \bigcup\limits_{i=1}^{K(a, t)}  \{  Y_i(a, t)  > n  \}   \right) 
  =   a \mu   \int_0^t n  e^{a (1 - \mu) (t - s)} \mathrm{B}\left(\dfrac{\lambda}{(1-\mu) s } + 1, n\right)    ds  . 
\end{equation}
To simplify the notation define $\tilde{\lambda} \equiv \lambda / (1 - \mu)$. 
Choosing $\epsilon > 0$ and omitting the exponential term, we obtain
\begin{equation}\label{eq:pn1}
(a\mu)^{-1} P_n(a, t)   
 \geq  \int_{\epsilon}^t  n  \mathrm{B}( \tilde{\lambda} /  s + 1, n)    ds 
 =  \int_{\epsilon}^t \dfrac{\Gamma(\tilde{\lambda}/  s + 1) \Gamma(n+1) }{ \Gamma( n + 1 + \tilde{\lambda}/  s ) }     ds . 
\end{equation}
On $\mathbb{R}^+$, $\Gamma(x)$ attains the minimum $\approx 0.885$ at $x \approx 1.46$. 
So we can bound $\Gamma( \tilde{\lambda}/   s + 1 )$ by $0.885$ or $\Gamma (\tilde{\lambda}/  t + 1)$ if $\tilde{\lambda}/ t  > 0.46$.  
For simplicity we henceforth assume $\tilde{\lambda}/ t  > 0.46$ and obtain
\begin{equation}
\begin{aligned}
 P_n(a, t)  & \; >   a \mu  \Gamma (\tilde{\lambda}/  t + 1) \int_\epsilon^t \dfrac{ \Gamma(n+1) }{ \Gamma( n + 1 + \tilde{\lambda}/ s ) } ds ,  \\
& \; \sim  a \mu  \Gamma (\tilde{\lambda}/  t + 1) \int_\epsilon^t \dfrac{1}{n^{\tilde{\lambda}/s } } ds \\
 &\; = a \mu  \Gamma (\tilde{\lambda} / t + 1)   \tilde{\lambda} \log n   \int^{  \tilde{\lambda} \log n/\epsilon }_{  \tilde{\lambda} \log n/t }  \dfrac{  e^{-x} }{x^2} dx .
\end{aligned}
\end{equation}
We can let $n$ go to infinity since $\tilde{\lambda}/s \geq \tilde{\lambda}/\epsilon$. 
The exponential integral is not an elementary function but can be bounded by (see \cite{abramowitz1964handbook})
\begin{equation}\label{eq:e1}
\dfrac{e^{-u} }{u^{n-1} (u + n)}   <     \int_{ u }^\infty   \dfrac{e^{-x}}{x^n} dx < \dfrac{e^{-u} }{u^{n-1} (u + n - 1)} \leq \dfrac{e^{-u}}{u^n} ,  \quad \quad  u > 0,  n = 1, 2, \dots  
\end{equation}
Hence,
\begin{equation}\label{eq:T2}
 \tilde{\lambda} \log n   \int^{  \tilde{\lambda} \log n/\epsilon }_{  \tilde{\lambda} \log n/t }  \dfrac{  e^{-x} }{x^2} dx  
 >  \dfrac{ t^2 n^{ - \tilde{\lambda} / t}   } {   \tilde{\lambda} \log n   + 2t   }  
- \dfrac{\epsilon^2 n^{ - \tilde{\lambda} / \epsilon}   }{   \tilde{\lambda} \log n     }.
\end{equation}
Note that in~\eqref{eq:pn1} we have omitted the integral from $0$ to $\epsilon$, which is of less interest to us.  
But using the inequality for beta function given in~\cite{cerone2007special} and~\eqref{eq:e1}, we can show that 
\begin{align*}
\int_0^\epsilon  n  e^{a (1 - \mu) (t - s)}  \mathrm{B}( \tilde{\lambda} /  s + 1, n)    ds  
 >  \dfrac{ e^{a(1 - \mu)(t - \epsilon)}     \epsilon^3  }{  (\tilde{\lambda} + \epsilon) ( \tilde{\lambda} \log n  + 3\epsilon ) }    n^{- \tilde{\lambda} / \epsilon}  ,  
\end{align*}
which grows at a slower rate (w.r.t. $n$) 
than~\eqref{eq:T2}. 
Since $\log n$ is a slowly varying function, for sufficiently large $n$, we have 
\begin{equation}\label{eq:def.L}
P_n(a, t)  >  \dfrac{ a \mu  t^2  \Gamma ( \tilde{\lambda} / t  + 1) }{  \tilde{\lambda} \log n  + 2t }  
  n^{- \tilde{\lambda} / t    } \equiv  L_n(a, t). 
\end{equation}
Finally, if we integrate over $a  \sim \text{exp}(\lambda)$ and recall the definition~\eqref{eq:def.nu}, we obtain 
\begin{equation}\label{eq:x2}
\mathbb{P}( X_2(t) > n) > P_n(t)  > \dfrac{ \mu \Gamma(\pconst + 1) }{ \pconst (  \pconst \log n  + 2) } n^{-\pconst}. 
\end{equation}

\paragraph*{Numerical examples}
We choose $a = 0.1, \mu = 0.2, \lambda = 10$ and simulate $10^6$ trajectories of $
X_2(a, t)$. 
The sample mean of $X_2(a, t)$ is $0.112$ at $t  = 4$  and $0.339$ at $t = 8$, which are equal to the theoretical values computed using~\eqref{eq:z22.expe}. 
The tail probabilities of $X_2(a, t)$ at $t = 8, 15$ are shown in Table~\ref{table:model2}.  
Recall that our estimate $P_n(a, t)$ defined in~\eqref{eq:pn} is a strictly lower bound for $ \mathbb{P} ( X_2(a, t) > n )$, and $L_n(a, t)$ defined in~\eqref{eq:def.L} is an asymptotic lower bound  for $P_n(a, t)$. 
Observe that in Table~\ref{table:model2}, both $P_n(a, t)$ and $L_n(a, t)$ can at least correctly estimate the order of the tail probabilities of $X_2(a,  t)$.  
In fact, $P_n(a, t)$ is very close to the sample average for large $n$, which is most likely due to the heavy tail of the distribution of $X_2(a, t)$. 
Furthermore, assuming the tail probability takes the form $n^x/\log n $, we estimate the exponent to be $-1.45$ for $t = 8$ and $-0.81$ for $t = 15$. 
They are very close to the theoretical values $-1.56$ for $t = 8$ and $-0.83$ for $t = 15$. 
Thus our estimate of the exponent, $\pconst = \lambda/(1-\mu)t$, is useful, although it tends to be slightly conservative. 

\begin{table}[h!]
\begin{center}
\begin{tabular}{cccccccccccc}
\toprule
  & $n$ &  5  &  10  & 15 &  20 &  25 &   50 & 100  &   200 \\ 
\hline 
\multirow{3}{*}{$t=8$ }  &  $\hat{\mathbb{P}}( X_2(t,a) > n ) \times 10^3$  &  5.38 &  1.49 & 0.69  & 0.40 & 0.28  & 0.07  & 0.02    & 0.009     \\  
   &  $ P_n(a, t) \times 10^3$  &  4.54 &  1.37  &  0.67  &  0.40  &  0.27  &  0.08  & 0.02  & 0.007   \\
      &  $ L_n(a, t) \times 10^3$  &  3.98  &  1.09 &  0.52  &  0.31  &  0.21  &  0.06  &  0.02  & 0.005   \\
 \hline 
      \multirow{3}{*}{$t=15$ }  &  $\hat{\mathbb{P}}( X_2(t, a) > n ) \times 10^3$  &  44.9 &  19.0  & 11.6  &  8.30  & 6.45  &  3.00  & 1.45 & 0.71   \\  
   &  $P_n(a, t) \times 10^3$  &  34.3  &  16.1  & 10.4  &  7.59  &  5.97  & 2.87  & 1.40  & 0.70     \\
      &  $L_n(a, t)  \times 10^3$  &  22.1  &  10.6  & 6.94  &  5.17  &  4.12  &  2.06  & 1.04  &  0.53  \\
 \bottomrule
\end{tabular}
\caption{Simulation of the model of primary and secondary tumors. 
The parameters are set as $a = 0.1, \mu = 0.2, \lambda = 10$. 
$\hat{\mathbb{P}}( X_2(a, t) > n )$ is the frequency in the $10^6$ simulated trajectories. 
$P_n(a, t)$ is defined in~\eqref{eq:pn} and computed by numerical integration. 
$L_n(a, t)$ is defined in~\eqref{eq:def.L}. 
} 
\label{table:model2}
\end{center}
\end{table}

\subsubsection{Tail probabilities of $X_k(a, t)$}
Such asymptotic analysis can be naturally extended to $X_k(a, t)$ for $k = 3, 4, \dots$. 
For example, when analyzing $X_3(a, t)$, we can treat the secondary tumor cells described by $X_2$ as primary tumor cells and apply our previous result. 
By both ~\eqref{eq:e1} and ~\eqref{eq:x2}, we obtain, for sufficiently large $n$,
\begin{align*}
\mathbb{P}( X_3(a, t) > n ) & \; >   a   \mu^2  \int_0^t \dfrac{ e^{a  (1 - \mu)(t - s)}  } { \tilde{\lambda} (  \tilde{\lambda} \log n + 2 s  ) } 
\Gamma( \tilde{\lambda} / s + 1) s^2 n^{- \tilde{\lambda} / s} ds \\
&\; >  \dfrac{ a  \mu^2 \Gamma( \pconst + 1)}{ \tilde{\lambda} (  \tilde{\lambda} \log n + 2 t  ) }   
\int_0^t s^2 n^{-\tilde{\lambda} / s} ds \\
&\; >  \dfrac{ a \mu^2 t^4 \Gamma(\pconst + 1)}{ \tilde{\lambda} (  \tilde{\lambda} \log n + 2 t  ) (  \tilde{\lambda} \log n + 4 t  )} n^{- \tilde{\lambda} / s}. 
\end{align*}
We can repeat this calculation and obtain the general expression of the tail probability of the $X_k(a, t)$. 
Assuming $(\tilde{\lambda} \log n + k t) \sim \tilde{\lambda} \log n$, we have  
\begin{align*}
\mathbb{P} ( X_k(a, t) > n )  >   C  a \left\{ \dfrac{  \mu (1 - \mu)^2 t^2 }{ \lambda^2 \log n }  \right\}^{k-1}    n^{- \pconst } , \quad \quad   n \rightarrow \infty,  \; k = 2, 3, \dots 
\end{align*}
where $C$ is a chosen constant.
This expression provides insight into the dynamics of the tumor cells. 
Firstly, the power law exponent $-\pconst$ is the same for all the tumor cells except the primary ones, but the growth rate of $X_k(a, t)$ is penalized by $(\log n)^{1-k}$. 
The exponent $\nu$ is equal to $1$ exactly when the expected value of the number of tumor cells explodes (recall~\eqref{eq:z1.moments} and~\eqref{eq:z22.expe}). 
Secondly, for small $t$, the tumor population is dominated by $X_1(a, t)$ and $X_2(a, t)$, but for large $t$, the cell populations $X_k(a, t)$ with large $k$ will eventually dominate. 
Lastly, given a moderate value of $t$, the value of $\mu$ will determine which of $X_1, X_2, \dots,$ dominates.
If $\mu$ is too small, then there will be no mutation to give rise to new subtypes. If $\mu$ is close to $1$, then no tumor subtypes will flourish since most divisions will not increase the total number of cells of that subtype.

\subsection{Towards general theory}
\textbf{Branching process with infinite type space}. We return to the modified G-C Model specified at the beginning of Section \ref{sec32}. Following the hypotheses of the model
and under the usual conditional independence assumptions, an ODE can
be written for the probability generating function) pgf of the distribution
of total cell count in all clones 
\beqn
F(s;a,t)=\expe{s^{X(a,t)}},s\in[0,1],
\eeqn
where $X(a,t)$ denotes the number of cells in the process started
by an ancestor of type $a$. The equation 
\begin{equation}
\frac{\partial F(s;a,t)}{\partial t}=-aF(s;a,t)+a[(1-\mu)F(s;a,t)^{2}+\mu F(s;a,t)\Phi(s;t)],\;t\geq0,\;a\geq0,\label{eq:6a}
\end{equation}
\begin{equation}
F(s;a,0)=s,\label{eq:6b}
\end{equation}
is analogous to the equation of the Coldman-Goldie model of clonal
resistance (\cite{Kimmel}), except that the pgf $\Phi(s,t)$
of the cell count of the clone started by a mutant of exponentially
distributed type is equal to 
\begin{equation}
\Phi(s,t)=\int\limits _{0}^{\infty}F(s;a',t)\cdot\lambda exp(-\lambda a')d\lambda,\label{eq:9}
\end{equation}
which follows from Hypothesis 2 of the modified G-C Model. Equation (\ref{eq:6a}) can be solved
and using Equation (\ref{eq:9}) compressed into a single integral
equation for $\Phi(s,t)$ (see the Appendix). It is also straightforward
to obtain
\beqn
M(a,t)=\expe{X(a,t)}=\frac{\partial F(s;a,t)}{\partial s}|_{s\uparrow1},
\eeqn
\beqn
\frac{\partial M(a,t)}{\partial t}=a(1-\mu)M(a,t)+a\mu\varphi(t),\label{eq:11}
\eeqn
where
\begin{equation}
\varphi(t)=\int\limits _{0}^{\infty}M(a',t)\lambda e^{-\lambda a'}da'\label{eq:12}
\end{equation}
is also equal to $\partial\Phi(s,t)/\partial s\,|_{s\uparrow1}$.
We can represent the solution of equation (\ref{eq:11}) using the variation
of constant formula
\begin{equation}
M(a,t)=g(t)+a\mu g(t)\stackrel{(t)}{*}\varphi(t),\label{ConvolEqu}
\end{equation}
where $\stackrel{(t)}{*}$ is the operator of convolution of functions
on $[0,\infty)$, and 
\begin{equation}
g(t)=e^{a(1-\mu)t}.\label{g(t)}
\end{equation}
Upon multiplying the equation by $\lambda e^{-\lambda a}$ and integrating
with respect to $a$ from $0$ to $\infty$, we obtain
\begin{equation} \label{integraleq}
\varphi(t)=f_{1}(t)+(\mu/\lambda)\,f_{2}(t)\stackrel{(t)}{*}\varphi(t),
\end{equation}
where
\beqn
f_{1}(t)=\int\limits _{0}^{\infty}g(t)\lambda e^{-\lambda a}da=\frac{1}{1-t(1-\mu)/\lambda},\;t\in[0,\lambda/(1-\mu)),
\eeqn
\beqn
f_{2}(t)=\int\limits _{0}^{\infty}ag(t)\lambda^{2}e^{-\lambda a}da=\left[\frac{1}{1-t(1-\mu)/\lambda}\right]^{2},\;t\in[0,\lambda/(1-\mu)).
\eeqn
Accordingly,
\bneqn
\varphi(t)=f_{1}(t)+f_{1}(t)\stackrel{(t)}{*}\sum_{i\ge1}(\mu/\lambda)^{i}f_{2}^{\stackrel{(t)}{*}i}(t),\label{phi}
\eneqn
where the infinite series of convolution powers converges uniformly
for $t$ in any closed subset of the interval $[0,\lambda/(1-\mu))$,
which can be proven by an argument akin to Picard iteration.

The expected value equations have solutions expressed as series of convolution
powers. They both explode at finite time $t=\lambda/(1-\mu)$. In addition, the function $\varphi(t)=\varphi(t,\lambda),$ is understood
as a function of two variables depends on $t/\lambda$ only, i.e.,
it has the following scaling property
\[
\varphi(t,k\lambda)=\varphi(t/k,\lambda),\;t\in[0,k\lambda/(1-\mu)).
\]

Numerical solutions based on the power series of (\ref{phi}) are depicted as thick continuous lines in Figs.~\ref{fig:fig2}, \ref{fig:fig3} and \ref{fig:fig4}, along with simulation averages. The relationship between these averages and the exact expected values $M(a,t)$ and $\varphi(t)$ is analogous to that of the toy model (Fig. 1 (D)).

\section{Conclusions}

This paper presents a series of models: a quasi-stochastic baseline model, a stochastic toy model, and a branching process model. The models propose a mechanism of generation heavy tail and ``explosive'' super-exponential growth of population of secondary tumors under very parsimonious assumptions. Our approach generates somewhat unexpected results without invoking new biological mechanisms. Of course, the finite-time ``explosions'' of expected values that we obtained will not occur in real word, in which cell proliferation rates ($a$) cannot be arbitrarily high, as required by the exponential distribution. This being said, the growth will still be accelerating if the exponential distributions of rates are truncated, in the sense that the expected values of cell counts in semi-log coordinates will be convex.

Equally important are the statistical and model building consequences. The analysis we carried out demonstrates that averages of empirical trajectories may be quite meaningless when building models of evolutionary phenomena such as cancer, in which heterogeneity plays a major role. Based on the toy model analysis, it is much more realistic to follow quantiles and deduce the growth law of the process from quantiles’ different growth exponents. Since the branching process model seems to behave very similarly to the toy model, this conclusion is likely to hold for it as well.

Mathematically, the analysis of the branching process model is quite preliminary. We can only conjecture the nature of the asymptotics of the Modified G-C model. We do not know which properties of the process persist if cell death is allowed. Finally, we do not know the mathematical structure of the Markov operator semigroup involved in such process. These questions certainly warrant further research.\\

\noindent \textbf{Acknowledgments} We thank Dinh Ngoc Khanh from the Applied Mathematics Department of the University of Alabama at Tuscaloosa for his help in Matlab.

\bibliographystyle{plain}
\bibliography{ref}

\section{Appendix}
\section*{Appendix - Solution of the baseline model}

We follow the approach of Iwata (\cite{Iwata}),
which involves a transport-type partial differential equation with
non-local boundary conditions, of the type considered among others
by \cite{Arino,Metz} and which can be used
to derive the distribution of the sizes of recurrent secondary tumors
shed by a growing primary. In the simplest cases, we can obtain closed-form
expressions. In the case in which the growth rates of the metastases
are exponentially distributed, we obtain expressions including incomplete
Gamma functions, which explode in the finite time.

\subsection*{Derivation of distribution density expression based on transport
equation}

\subsubsection*{Case 1. Primary and metastatic tumors grow at the same rate $a=b$}

\noindent From equation (\ref{Baseline}) and the exponential growth
rate hypothesis $g(x)=ax$, we derive the following transport equation
\begin{eqnarray*}
\frac{\partial\rho}{\partial t}+ax\frac{\partial\rho}{\partial x} & = & -a\rho.
\end{eqnarray*}

\noindent Equivalently, if $x\neq0$,

\begin{eqnarray*}
\frac{\partial\rho}{\partial x}+\frac{1}{ax}\frac{\partial\rho}{\partial t} & = & -\frac{\rho}{x}.\label{E4}
\end{eqnarray*}
Assuming
$x$ as the independent variable, we apply the method of characteristics
\begin{eqnarray*}
\tilde{\rho}(x)=\rho(x,\tau(x)),
\end{eqnarray*}
where $\tilde{\rho}$ denotes the distribution density ($\rho$) parameterized
along characteristics. Integrating the equation $d\tau/dx=(ax)^{-1}$
from 1 to $x$, we obtain
\begin{eqnarray}
\tau(x)-\tau(1)=\frac{1}{a}\ln x,\label{E8}
\end{eqnarray}
which leads to the solution of the form  
$
\rho(x,\tau(x))=\rho(1,\tau(1))/x.
$
Considering equation (\ref{E8}) and writing $t=\tau(x)$, we obtain
\begin{eqnarray}
\rho(x,t)=\frac{\rho(1,t-\frac{1}{a}\ln|x|)}{x},\label{E13}
\end{eqnarray}
which implies that $\rho(x,t)=0$, $x>e^{at}$.

\indent We assume that at time $t=0$ no metastatic tumor exists. Therefore, the initial condition is
\begin{eqnarray}
\rho(x,0)=0.\label{initial}
\end{eqnarray}
The boundary condition at $x=1$ has the  non-local form given in (\ref{nonlocal}).
Equation (\ref{nonlocal}) indicates that the number of metastatic
single cells newly created per unit time at time $t$ (the left-hand
side term) is the sum of the total rate of occurrence of
metastases due to metastatic tumors and the primary tumor (corresponding
to the first and second terms of the right-hand side) {\cite{Iwata}.
$x_{p}(t)$ represents the number of cells in the primary tumor at
time $t$, which is given by the solution of 
\begin{eqnarray}
\frac{dx_{p}}{dt}=g(x_{p}),\ \ \ \ x_{p}(0)=1\label{dx_p}.
\end{eqnarray}

\noindent Using $G(x)=ax$ in equation (\ref{dx_p}) , we obtain the number of cells in the
primary tumor as a function of time:
\begin{eqnarray}
x_{p}(t)=e^{at}.\label{E18}
\end{eqnarray}
\noindent Denoting $\rho_{1}(t)=\rho(1,t)$ and substituting equations
(\ref{E13}) and (\ref{E18}) into equation (\ref{nonlocal}) yields
\begin{eqnarray}
a\rho_{1}(t)=\int\limits _{1}^{e^{at}}mx^{\alpha}\rho_{1}(t-\frac{1}{a}\ln|x|)\frac{1}{x}dx+me^{a\alpha t}.\label{E20}
\end{eqnarray}
Following a the change of variables $x=e^{a(t-u)}$, $dx=-ax\ du,$ (\ref{E20}) can be reexpressed as
\begin{eqnarray*}
a\rho_{1}(t)=ame^{at}*\rho_{1}(t),
\end{eqnarray*}
where 
\begin{eqnarray*}
(f*g)(t)=\int\limits _{0}^{t}f(t-\tau)g(\tau)d\tau.
\end{eqnarray*}
Passing to Laplace transforms $\hat{\rho}{_1}(s)$, we obtain
\begin{eqnarray}
a\hat{\rho}_{1}(s)=a\frac{m\hat{\rho}_{1}(s)}{s-a\alpha}+\frac{m}{s-a\alpha},
\end{eqnarray}
and
\begin{eqnarray*}
\hat{\rho}_{1}(s)=\frac{\frac{m}{a}}{s-(a\alpha+m)},
\end{eqnarray*}
which lead to
\begin{eqnarray*}
\rho_{1}(t)=\frac{m}{a}e^{(a\alpha+m)t}.
\end{eqnarray*}
Using equation (\ref{E13}), we obtain
\begin{eqnarray}
\rho(x,t)=\frac{\frac{m}{a}e^{(a\alpha+m)(t-\frac{1}{a}\ln x)}}{x}=\frac{m}{a}e^{(a\alpha+m)t}x^{-(\alpha+\frac{m}{a}+1)},\ \ \ \ x\leq e^{at}.
\end{eqnarray}

\noindent Let $G(x)$ be the number of migrant clones which have more
than $x$ cells at time $t$, 
\begin{eqnarray}
G(x) & = & \int\limits _{x}^{e^{at}}\rho(\xi,t)d\xi
 \\
 & = & \frac{m}{a\alpha+m}(e^{(a\alpha+m)t}x^{-(\alpha+\frac{m}{a})}-1)\ \ \ \ x<e^{at}.\label{G}
\end{eqnarray}

\subsubsection*{Case 2. Metastatic tumors grow at a different rate, $a\protect\neq b$}

Equation (\ref{G}) may be extended to include growth advantage, the newly seeded tumor having growth rate, which
may be higher or lower than the growth rate of the parent tumor. Using equation (\ref{nonlocal}) and equation for the number of
cells in the primary tumor 
\begin{eqnarray*}
x_{p}(t)=e^{bt},
\end{eqnarray*}
we obtain 
\begin{eqnarray*}
a\rho_{1}(t)=\int\limits _{1}^{e^{at}}mx^{\alpha}\rho_{1}(t-\frac{1}{a}\ln|x|)\frac{1}{x}dx+me^{\alpha bt}.
\end{eqnarray*}
or, after a change of variables, 
\begin{eqnarray*}
a\rho_{1}(t)=a\int\limits _{1}^{t}me^{a\alpha(t-u)}\rho_{1}(u)du+me^{\alpha bt}.
\end{eqnarray*}
Again, we use the Laplace transform to obtain 
\begin{eqnarray}
\hat{\rho}_{1}(s)=\frac{m}{a}\frac{s-a\alpha}{(s-\alpha b)(s-a\alpha-m)}.\label{rholambda}
\end{eqnarray}
The inverse Laplace transform yields
\begin{equation}
\rho_{1}(t)=\frac{m}{a(a\alpha+m-\alpha b)}((a\alpha-\alpha b)e^{\alpha bt}+me^{(a\alpha+m)t}.
\end{equation}
Using equation (\ref{E13}) results, for $x\le e^{at}$, in
\begin{eqnarray}
\rho(x,t) & = & \frac{(\frac{m}{a(a\alpha+m-\alpha b)}((a\alpha-\alpha b)e^{\alpha b(t-\frac{1}{a}\ln|x|)}+me^{(a\alpha+m)(t-\frac{1}{a}\ln|x|)}))}{x}\\
 & = & \frac{m}{a(a\alpha+m-\alpha b)}((a\alpha-\alpha b)e^{\alpha bt}x^{-(\frac{\alpha b}{a}+1)}+me^{(a\alpha+m)t}x^{-(\alpha+\frac{m}{a}+1)}),\nonumber 
\end{eqnarray}

\noindent Correspondingly, $G(x)=G(x;a,b)$, which is
the number of migrant clones with more than $x$ cells, is given for $x\in[1,e^{at}]$ as
\begin{eqnarray}
G(x;a,b)= \frac{m}{a\alpha+m-\alpha b}(\frac{a-b}{b}(e^{\alpha bt}x^{-\frac{\alpha b}{a}}-1)+\frac{m}{a\alpha+m}(e^{(a\alpha+m)t}x^{-\alpha-\frac{m}{a}}-1)).\nonumber 
\nonumber 
\end{eqnarray}
For $x>e^{at}$, $G(x)=0$.

\subsubsection*{Case 3. Metastatic growth rate has exponential distribution with
parameter $\lambda$.}

\noindent In this case, we obtain 
\begin{equation*}
\tilde{G}(x,b)=\int\limits _{0}^{\infty}G(x;a,b)\lambda e^{-\lambda a}da.
\end{equation*}
Taking into account that $G(x;a,b)=0$ for $x>e^{at}$ or equivalently
for $a<\frac{1}{t}\ln x$ we obtain 
\begin{equation*}
\tilde{G}(x,b)=\int\limits _{\frac{1}{t}\ln x}^{\infty}G(x;a,b)\lambda e^{-\lambda a}da.
\end{equation*}
In general this integral seems analytically intractable. However,
the special case $x=1$ can be expressed in the terms of the so-called
incomplete gamma functions. $\tilde{G}(1,b)$ is important since it is equal
to the total metastasis load at time t. Recall the incomplete Gamma function (IGF) is defined for positive $x$ and complex $a$ as follows
\[
\Gamma(c,x)=\int_{x}^{\infty}e^{-t}t^{c-1}dt.
\]
The expression for $G(1;a,b)$ has
the form 
\[
G(1;a,b)=\frac{m}{a\alpha+m-\alpha b}(\frac{a-b}{b}(e^{\alpha bt}-1)+\frac{m}{a\alpha+m}(e^{(a\alpha+m)t}-1)).
\]

\noindent In the above equation we can distinguish following terms,
which after multiplication by $\lambda e^{-\lambda a}$ and integration
from $\frac{1}{t}\ln x$ (which in this case is equal to zero) to infinity
give us the solution in the form of incomplete Gamma functions and elementary
functions. 
\begin{itemize}
\item $\frac{m}{a\alpha+m-b\alpha}\frac{a-b}{b}e^{b\alpha t}$ which leads
to $\int_{0}^{\infty}\frac{me^{b\alpha t}}{\alpha b}\lambda e^{-\lambda a}da-\int_{0}^{\infty}\frac{m^{2}e^{b\alpha t}}{\alpha^{2}b(a+\frac{m}{\alpha}-b)}\lambda e^{-\lambda a}da$,
where solution of the first term is elementary ($\frac{1}{\lambda}$)
and the solution of second term leads to the incomplete Gamma function
with parameters included in Table \ref{Tab1}, term number 1. 
\item (-) $\frac{m}{a\alpha+m-b\alpha}\frac{a-b}{b}$ which leads to $-\int_{0}^{\infty}\frac{m}{\alpha b}\lambda e^{-\lambda a}da+\int_{0}^{\infty}\frac{m^{2}}{\alpha^{2}b(a+\frac{m}{\alpha}-b)}\lambda e^{-\lambda a}da$,
where solution of the first term is elementary ($\frac{1}{\lambda}$)
and the solution of second term leads to the incomplete Gamma function
with parameters included in Table \ref{Tab1}, term number 2. 
\item $\frac{m}{a\alpha+m-b\alpha}\frac{m}{a\alpha+m}e^{(a\alpha+m)t}$
which leads to $\int_{0}^{\infty}\frac{m^{2}e^{(a\alpha+m)t}}{\alpha^{2}b(a+\frac{m}{\alpha}-b)}\lambda e^{-\lambda a}da-\frac{m^{2}e^{(a\alpha+m)t}}{\alpha^{2}b(a+\frac{m}{\alpha})}\lambda e^{-\lambda a}da$
, where solutions of the both terms lead to the incomplete Gamma functions
with parameters included in Table \ref{Tab1}, term number 3 and term
number 4. 
\item (-) $\frac{m}{a\alpha+m-b\alpha}\frac{m}{a\alpha+m}$ which leads to
$-\int_{0}^{\infty}\frac{m^{2}}{\alpha^{2}b(a+\frac{m}{\alpha})}\lambda e^{-\lambda a}da+\frac{m^{2}}{\alpha^{2}b(a+\frac{m}{\alpha})}\lambda e^{-\lambda a}da$
, where solutions of the both terms lead to Incomplete Gamma Functions
with parameters included in Table \ref{Tab1}, term number 5 and term
number 6. 
\end{itemize}
\begin{table}[!t]
\centering \caption{Elements of the equation. Terms with Incomplete Gamma Function have
the form of: $\text{Coefficient}\cdot e^{CD}\Gamma[0,\,CD]$\label{Tab1}}

{%
\begin{tabular}{lcccccc}
 &  &  &  &  &  & \tabularnewline
\midrule 
Term  & 1  & 2  & 3  & 4  & 5  & 6 \tabularnewline
\midrule 
Coefficient  & $-\frac{\lambda m^{2}e^{b\alpha t}}{\alpha^{2}b}$  & $\frac{\lambda m^{2}}{\alpha^{2}b}$  & $-\frac{\lambda m^{2}e^{mt}}{\alpha^{2}b}$  & $\frac{\lambda m^{2}e^{mt}}{\alpha^{2}b}$  & $\frac{\lambda m^{2}}{\alpha^{2}b}$  & $-\frac{\lambda m^{2}}{\alpha^{2}b}$ \tabularnewline
C  & $\frac{m}{\alpha}-b$  & $\frac{m}{\alpha}-b$  & $\frac{m}{\alpha}-b$  & $\frac{m}{\alpha}$  & $\frac{m}{\alpha}-b$  & $\frac{m}{\alpha}$ \tabularnewline
D  & $\lambda$  & $\lambda$  & $\lambda-\alpha t$  & $\lambda-\alpha t$  & $\lambda$  & $\lambda$ \tabularnewline
\end{tabular}} 
\end{table}

\noindent Let us first consider the the following term present in
$G(1;a,b)$, $\frac{m}{a\alpha+m-b\alpha}\frac{m}{a\alpha+m}$, which
has to be multiplied by $\lambda e^{-\lambda a}$ and integrated from
$0$ to infinity. We obtain 
\[
-\int_{0}^{\infty}\frac{m}{a\alpha+m-b\alpha}\frac{m}{a\alpha+m}\lambda e^{-\lambda a}da=\frac{\lambda m^{2}}{\alpha^{2}b}\int_{0}^{\infty}\frac{e^{-\lambda a}}{a+\frac{m}{\alpha}-b}da-\frac{\lambda m^{2}}{\alpha^{2}b}\int_{0}^{\infty}\frac{e^{-\lambda a}}{a+\frac{m}{\alpha}}da
\]
\[
=\frac{\lambda m^{2}}{\alpha^{2}b}\int_{\frac{m}{\alpha}-b}^{\infty}\frac{e^{-\lambda a}}{u}du-\frac{\lambda m^{2}}{\alpha^{2}b}\int_{\frac{m}{\alpha}}^{\infty}\frac{e^{-\lambda a}}{w}dw=\frac{\lambda m^{2}}{\alpha^{2}b}\int_{(\frac{m}{\alpha}-b)\lambda}^{\infty}\frac{1}{v}e^{-v}dv-\frac{\lambda m^{2}}{\alpha^{2}b}\int_{\frac{\lambda m}{\alpha}}^{\infty}\frac{1}{z}e^{-z}dz
\]
\[
=\frac{\lambda m^{2}}{\alpha^{2}b}(\Gamma(0,\ (\frac{m}{\alpha}-b)\lambda)-\Gamma(0,\ \frac{\lambda m}{\alpha})),
\]

\noindent where the first equality follows from substitution $u=a+\frac{m}{\alpha}-b$
and $w=a+\frac{m}{\alpha}$, while the second follows by $v=\lambda u$ and
$z=\lambda w$. Integration of the remaining terms also follows analogously. The final form of the equation is 
\begin{eqnarray}
\tilde{G}(1;b) & = & \frac{m(e^{b\alpha t}-1)}{\alpha b}+(2-e^{b\alpha t})\frac{\lambda m^{2}}{\alpha^{2}b}e^{\lambda(\frac{m}{\alpha}-b)}\Gamma(0,\ \lambda(\frac{m}{\alpha}-b))\nonumber \\
 & + & \frac{\lambda m^{2}}{\alpha^{2}b}(-e^{mt}(e^{(\lambda-\alpha t)(\frac{m}{\alpha}-b)}\Gamma(0,\ (\lambda-\alpha t)(\frac{m}{\alpha}-b))-e^{(\lambda-\alpha t)\frac{m}{\alpha}}\Gamma(0,\ (\lambda-\alpha t)\frac{m}{\alpha}))\nonumber \\
 & - & e^{\frac{\lambda m}{\alpha}}\Gamma(0,\ \frac{\lambda m}{\alpha}))
\end{eqnarray}
for $b<m/\alpha$ and $\lambda>\alpha t$.

\captionsetup[figure]{font=footnotesize}
\section*{Figures}
\begin{figure}[h!]
\begin{center}
\includegraphics[width=0.9\linewidth]{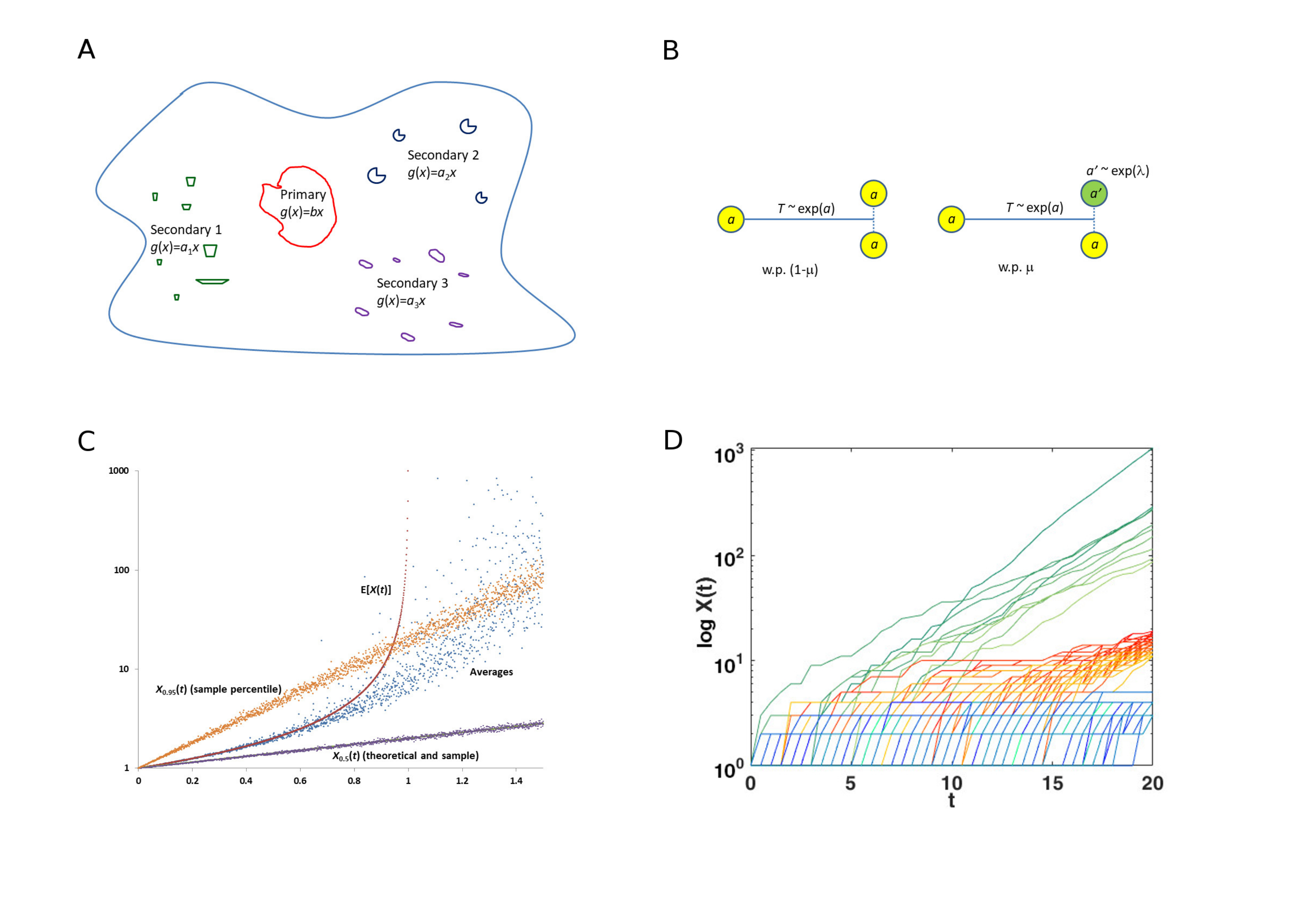} 
\vspace{-1cm}
\caption{
(A)  Ideogram representation of the baseline process. Within the `tumor field', primary tumor is growing exponentially at rate $b$, and then sheds secondary tumors, which may shed further secondary tumors. Secondary tumors grow at rates generally different from that of the primary tumor.   
(B)  Hypotheses underlying the Modified Goldie-Coldman model.
(C)  Toy model with $\lambda=1$: Expected value $\expe{X(t)}$ of the process (which explodes at $t=1$), averages of 1000 realizations of $X(t)$, and 0.5 and 0.95 quantiles of $X(t)$, all in semi-logarithmic scale.
(D)  Summary of 10,000 simulated trajectories of the Modified G-C process with parameters $\mu=0.5$, $a=0.01$, $\lambda=10$. Depicted are only realizations of the process ranking 1-10 (green), 51-100 (red), and 301-400 (blue) at time $t=20$. 
}\label{fig:fig1}
\end{center}
\end{figure}

\begin{figure}[h!]
\begin{center}
\includegraphics[width=0.8\linewidth]{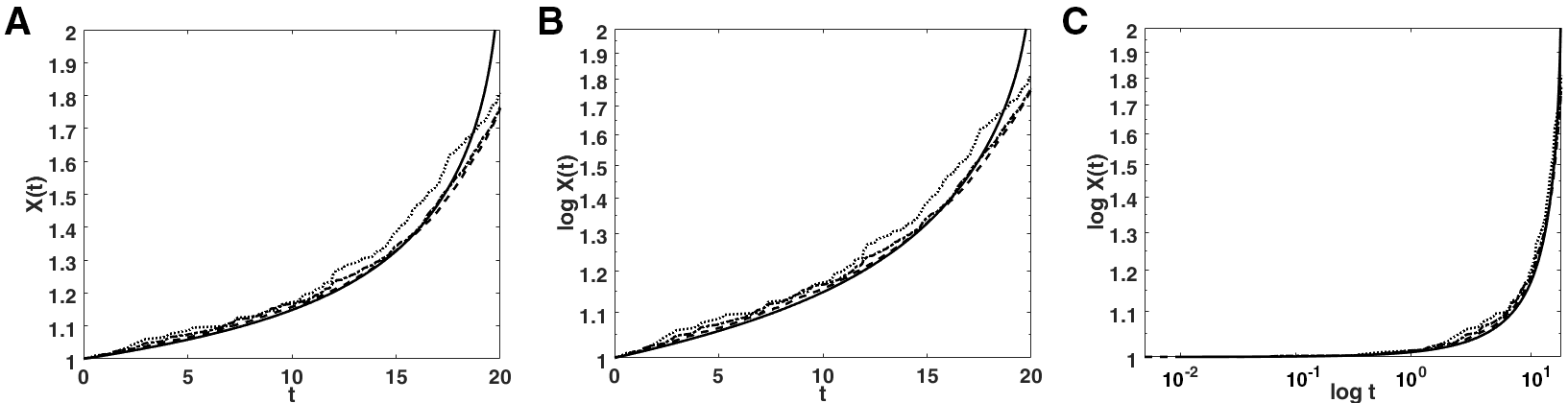}\\
\caption{Averages of the simulated trajectories of the Modified G-C process with parameters $\mu=0.5$, $a=0.01$, $\lambda=10$, based on 200, 1000, and 10,000 trajectories (dotted, dashed-dotted, and dashed lines, respectively), with the expectation $M(a,t)$, computed by numerically solving the integral equation (\ref{integraleq}) for $\varphi(t)$ and using expression \ref{ConvolEqu}. (A) $t$ and $X(t)$ in linear scale, (B) $t$ in linear and $X(t)$ in logarithmic scale, (C) $t$ and $X(t)$ in logarithmic scale. }\label{fig:fig2}
\end{center}
\end{figure}

\begin{figure}[h!]
\begin{center}
\includegraphics[width=0.8\linewidth]{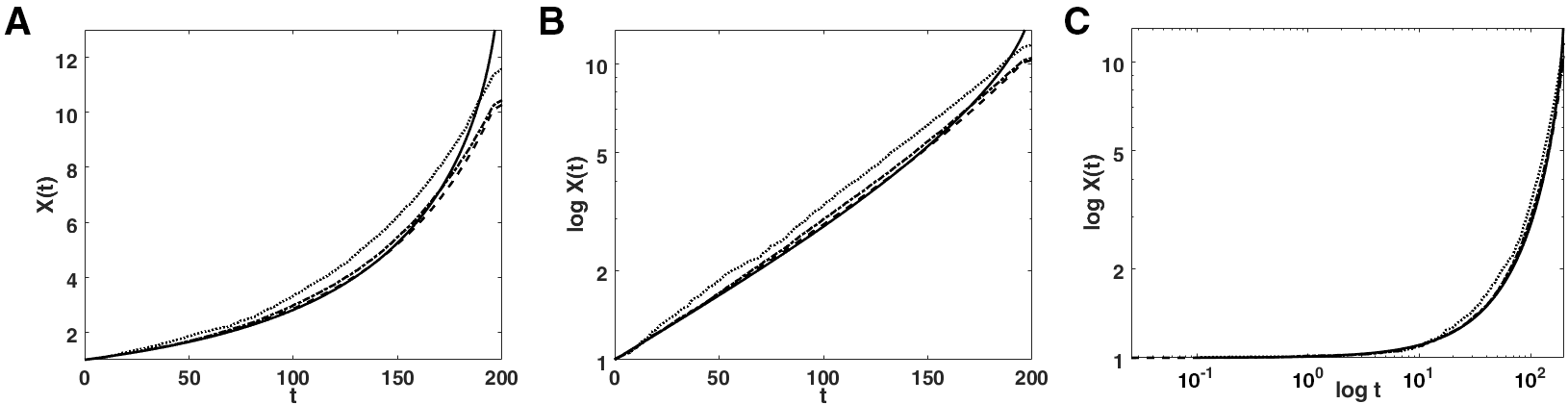}\\
\caption{Averages of the simulated trajectories of the Modified G-C process with parameters\,$\mu=0.5$, $a=0.01$, $\lambda=100$. Details as in Fig.~\ref{fig:fig2}.}\label{fig:fig3}
\end{center}
\end{figure}

\begin{figure}[h!]
\begin{center}
\includegraphics[width=0.8\linewidth]{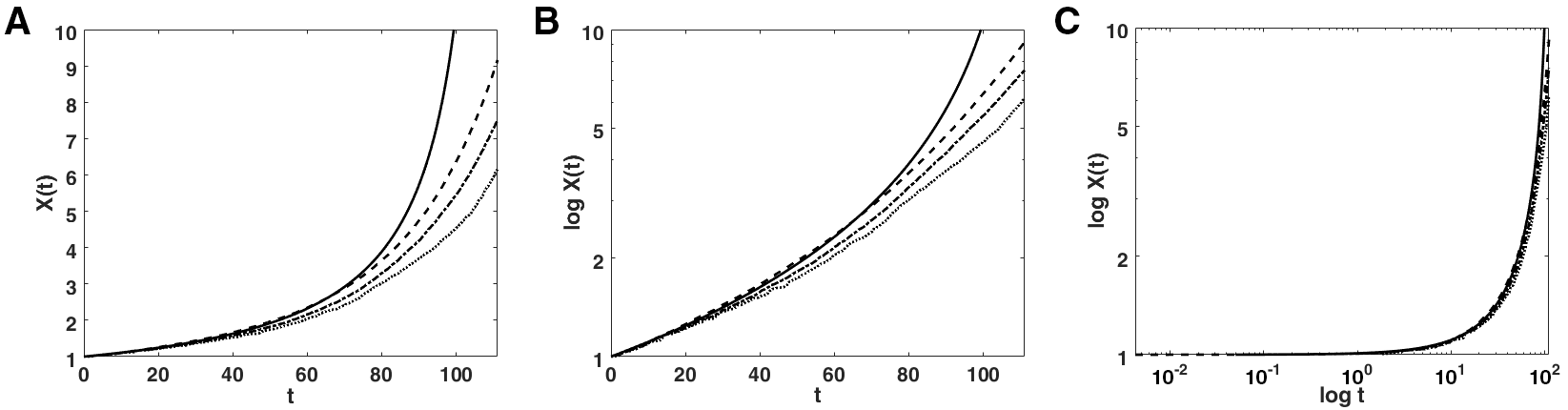}\\
\caption{ Averages of the simulated trajectories of the Modified G-C process starting from a cell with randomly selected parameter of lifetime distribution, with parameters \,$\mu=0.1$, $\lambda=10$0, based on 200, 1000, and 10,000 trajectories (dotted, dashed-dotted, and dashed lines, respectively), with the expectation $\varphi(t)$, computed by numerically solving the integral equation (\ref{integraleq}). 
(A) $t$ and $X(t)$ in linear scale, 
(B) $t$ in linear and $X(t)$ in logarithmic scale, (C) $t$ and $X(t)$ in logarithmic scale.
}\label{fig:fig4}
\end{center}
\end{figure}

\begin{figure}[h!]
\begin{center}
\includegraphics[width=0.6\linewidth]{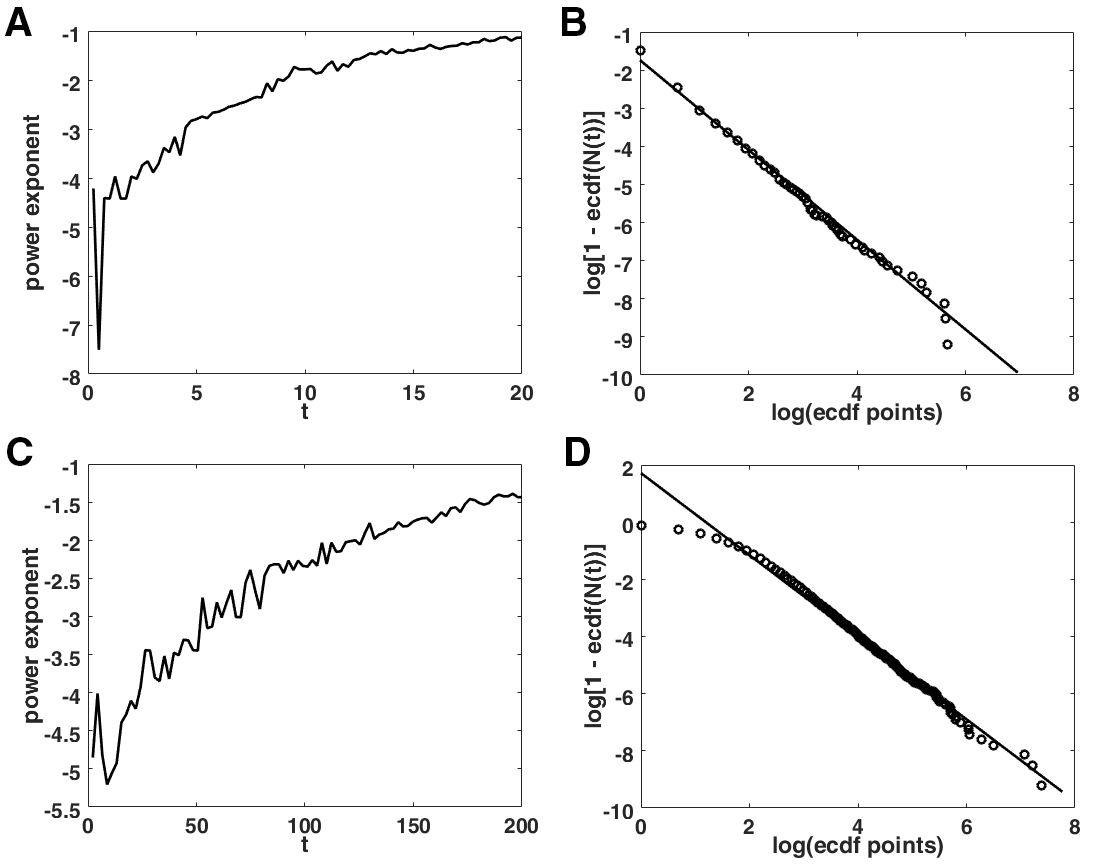}\\
\caption{Simulated tail behavior of the Modified G-C process with parameters \,$\mu=0.5$, $a=0.01$, $\lambda=10$ (A, B)  and $\lambda=100$ (C, D). 
Depicted are estimated power exponents of the tail of $X(a,t)$, $t\in[0,\lambda/(1-\mu)]$(left) and examples of empirical tails at the expected value explosion times, in log-log coordinates, approximated by a straight lines (right).
}\label{fig:fig5}
\end{center}
\end{figure}

\end{document}